\documentclass[aip,jap,amsmath,amssymb,reprint,twocolumn]{revtex4-1}

\usepackage[utf8]{inputenc}
\usepackage{graphicx}% Include figure files
\usepackage{dcolumn}% Align table columns on decimal point
\usepackage{bm}% bold math
\usepackage{upquote}
\usepackage{multirow}
\usepackage{tikz}
\usetikzlibrary{arrows}

\begin{document}

\newcommand {\bc}{\begin{center}}
\newcommand {\ec}{\end{center}}
\newcommand {\be}{\begin{equation}}
\newcommand {\ee}{\end{equation}}
\newcommand {\ben}{\begin{equation*}}
\newcommand {\een}{\end{equation*}}
\newcommand {\beq}{\begin{eqnarray}}
\newcommand {\eeq}{\end{eqnarray}}
\newcommand {\beqn}{\begin{eqnarray*}}
\newcommand {\eeqn}{\end{eqnarray*}}
\newcommand {\nn}{\nonumber}
\newcommand {\bpm}{\begin{pmatrix}}
\newcommand {\epm}{\end{pmatrix}}
\newcommand {\bi}{\begin{itemize}}
\newcommand {\ei}{\end{itemize}}
\newcommand {\upa}{\uparrow}
\newcommand {\doa}{\downarrow}
\newcommand {\Upa}{\Uparrow}
\newcommand {\Doa}{\Downarrow}
\newcommand{\bibitemShut}[1]{}
\providecommand{\e}[1]{\ensuremath{\times 10^{#1}}}

\preprint{AIP/123-QED}

\title[]{Characterization of the nitrogen split interstitial defect in wurtzite aluminum nitride using density functional theory}

\author{A. Sz\'all\'as}
 \email{szallas.attila@wigner.mta.hu}
 \affiliation{Institute for Solid State Physics and Optics, Wigner Research Centre for Physics,\\ Hungarian Academy of Sciences, P.O. Box 49, H-1525 Budapest, Hungary}
\author{K. Sz\'asz}
 \affiliation{Institute for Solid State Physics and Optics, Wigner Research Centre for Physics,\\ Hungarian Academy of Sciences, P.O. Box 49, H-1525 Budapest, Hungary}
 \affiliation{Institute of Physics, E\"otv\"os University,\\ P\'azm\'any P\'eter s\'et\'any 1/A, H-1117 Budapest, Hungary}
\author{X. T. Trinh}
\author{N. T. Son}
\author{E. Janz\'en}
\affiliation{Department of Physics, Chemistry and Biology, Link\"oping University,\\ SE-581 83 Link\"oping, Sweden}
\author{A. Gali}
 \email{gali.adam@wigner.mta.hu}
 \affiliation{Institute for Solid State Physics and Optics, Wigner Research Centre for Physics,\\ Hungarian Academy of Sciences, P.O. Box 49, H-1525 Budapest, Hungary}
 \affiliation{Department of Atomic Physics, Budapest University of Technology and Economics,\\ Budafoki \'ut 8., H-1111 Budapest, Hungary}

\date{\today}% It is always \today, today,

\begin{abstract}
We carried out Heyd-Scuseria-Ernzerhof hybrid density functional theory plane wave supercell calculations in wurtzite aluminum nitride in order to characterize the geometry, formation energies, transition levels and hyperfine tensors of the nitrogen split interstitial defect. The calculated hyperfine tensors may provide useful fingerprint of this defect for electron paramagnetic resonance measurement.
\end{abstract}

\maketitle

\section{Introduction}

The wurtzite phase of aluminum nitride (w-AlN) is a technical ceramics and wide band gap semiconductor that has many applications in industry. Like for other group III nitrides, such as gallium nitride (GaN) and indium nitride (InN), probably the most important applications are related to optoelectronics \cite{Walle_Neugebauer_JAP_2004}. One of the well known examples for optoelectronic application of group III nitrides is the Blu-ray disc optical storage technology, which mostly uses GaN crystal as an active medium in the solid state laser operating at 405 nm for reading and writing the disc. Generally, the green, blue, and ultraviolet lasers and light-emitting diodes mostly use group III nitrides and their alloys \cite{Nakamura_SSC_1997}. AlN and AlGaN alloys emit and adsorb light with wavelength from 200 to 300 nm. Laser diode, which emit at 210 nm, has already been succesfully fabricated using AlN \cite{taniyasu2006aluminium}. These materials are ideal for the development of chip-scale UV light sources and sensors. Furthermore, group III nitrides are used not in lasers and sensors only, but for example, AlGaN/GaN heterojunction field effect transistors have attracted great attention among high-electron-mobility transistors (HEMT), due to their high-power performance. The useful properties that make these materials inevitable in the above mentioned applications originate from their mechanical stability and electronic properties, especially from the wide band gap they exhibit.\\
\indent Defects may introduce deep levels in the band gap and can significantly alter optical properties of semiconductors. Our knowledge on basic defects in nitrides is far from complete understanding despite of their importance. Among the group III nitrides, GaN is the most thoroughly investigated one, yet very few intrinsic defects of this compound have been identified up to now, such as Ga interstitial \cite{Chow_PRL_2000} and Ga monovacancy \cite{Saarinen_PRB_2001}. Recently, the nitrogen (N) split interstitial defect [$\text{(N-N)}_\text{N}$] has been identified \cite{Bardeleben_PRL_2012} in GaN. To produce defects in a controlable fashion, an n-type GaN bulk crystal was grown by hybrid vapor phase epitaxy and irradiated by high energy electrons, high energy protons, and swift heavy Si ions. A single defect with $C_{1h}$ symmetry was detected for proton fluences above $10^{16} /$cm$^2$, using electron paramagnetic resonance (EPR) spectroscopy. As the EPR measurement with simultaneous photoexcitation showed that the defect pins the Fermi level at $E_C - 1.0$ eV, this point defect has been assigned as an N interstitial. Comparing the measured hyperfine structure with the result of ab initio model calculation, the N interstitial has been identified as a neutral N split interstitial defect with the total spin of $S=1/2$.\\
\indent These results and the above mentioned existing and potential applications, especially that are related to the AlGaN compound, motivated a research to identify the N split interstitial defect in the less known AlN as well. The wurtzite phase of the AlN has a wide band gap of $6.03 - 6.12$\,eV \cite{Guo_JJAP_1994, Li_APL_2003} (from room temperature to 0 K) that is even wider than for GaN and InN, and thus exhibit a great potential for applications for deep ultraviolet optoelectronics \cite{Berger_1996}. We characterized this defect in w-AlN using highly accurate density functional theory calculations, in order to foster the defect identification, especially by future EPR measurements. We calculated the electronic band structure, formation energies, transition levels, and hyperfine properties for this defect in w-AlN.

\section{Method of the calculation}
The calculations were performed using the 5.3.3 version of the Vienna Ab Initio Simulation Package (VASP) \cite{Kresse_Hafner_PRB_1994,Kresse_Furthmueller_PRB_1996}. For accurate calculation of the spin density close to the nuclei we used all-electron projector augmented wave method (PAW) \cite{PhysRevB.50.17953,PhysRevB.59.1758} and plane wave basis set. The standard PAW-projectors was applied that are available in the VASP package. The plane wave cut-off was set up to 420 eV for geometry optimization that was sufficient to achieve convergence for the structure. However, for convergent spin density and hyperfine constants a higher plane wave cut-off, 600 eV was sufficient to apply. We used 432-atom supercell for modelling a single defect, which is large enough to minimize the magnitude of the finite size effects. For these large supercells the $\Gamma$-point sampling sufficed. Our calculations are strictly valid at the limit of T $=0$\,K as we basically applied the Born-Oppenheimer approximation. We used the Heyd-Scuseria-Ernzerhof (HSE) hybrid functional \cite{Heyd_Scuseria_Ernzerhof_JCP_2003} that can provide more accurate results for the hyperfine constants concerning the point defects in semiconductor \cite{PhysRevB.88.075202}. The original HSE hybrid functional (HSE06) applies the mixing parameter $a=\frac{1}{4}$ and screening parameter $\omega=0.2$ 1/\AA. Using these parameters, the band gap ($E_\text{gap}$) is 5.65 eV \cite{Ivady_PRB_2014} for bulk w-AlN crystal that still underestimates the extrapolated $6.12$ eV at $0$ K, comes from optical absorption measurement \cite{Guo_JJAP_1994} or photoluminescence measurement \cite{Li_APL_2003}. Considering the suggestion of Ref.~\onlinecite{Ivady_PRB_2014}, we kept $a=\frac{1}{4}$ fixed and modified the $\omega$ to improve the accuracy of the $E_\text{gap}$ in the calculations. We found a linear scaling of $E_\text{gap}$ with $\omega$, such as $E_\text{gap}(\omega) = 6.35 \text{ eV} - 3.8(0) \cdot \omega \text{ \AA} \text{ eV}$. The $E_\text{gap}$ of $6.12$ eV is reached around $\omega = 0.061$ 1/\AA, and we used this $\omega$ value for all the calculations. In our optimized geometries all the forces were smaller than 1\e{-2} eV/\AA.\\
\indent We calculated the formation energy of the different charge states in both Al-rich and N-rich cases. We defined the formation energy as \cite{Walle_Neugebauer_JAP_2004}:
\begin{eqnarray*}
E^f[\text{X}^q]=E_\text{tot}[\text{X}^q]-E_\text{tot}[\text{AlN},\text{bulk}] \\
-\mu_N + q[E_\text{F} + E_\text{v}] + \Delta V,
\end{eqnarray*}
where $E_\text{tot}[\text{X}]$ is the total energy obtained from a supercell calculation with defect X in the cell, and $E_\text{tot}[\text{AlN},\text{bulk}]$ is the total energy for the equivalent defect free supercell. $\mu_N$ is the chemical potential of the nitrogen either in Al-rich or in N-rich cases. The charge state of the defect is $q$. $E_\text{F}$ is the Fermi energy and $E_\text{v}$ is the reference point, in our case the valence band maximum (VBM) of the pristine crystal. $\Delta V$ is the charge correction term that corrects the effect of the electrostatic potential between the charged defect and its periodic images. We considered the charge correction scheme described by Freysoldt, Neugebauer and Van de Walle \cite{Freysoldt_PRL_2009, Freysoldt_PSS_2010}. This correction scheme supports defect charge distribution for every shape of the supercell. The obtained charge corrections are $0.16$ eV and $0.59$\,eV for defects with $|q| = 1$ and for $|q| = 2$, respectively, following approximately the $\sim q^2$ dependence of the charge correction in agreement with the Makov-Payne correction scheme.\\
\indent For the formation energy, we calculated the total energy of the bulk Al fcc structure and the N$_2$ molecule. From these calculation we could estimate the enthalpy of formation for AlN. We obtained $\Delta H_\text{f}=-3.18$ eV for the enthalpy of formation that agrees well with the experimental measurements [$-3.13$\,eV\,\cite{Tb_Chemiker_Physiker},$-3.28$\,eV\,\cite{Group_III_Nitrides}]. The transition level between the charge states $q$ and $q'$ is the position of the Fermi level $E_F$, where $E^f[\text{X}^q](E_\text{F}) = E^f[\text{X}^{q'}](E_\text{F})$ is satisfied, which is denoted as $(q/q')$ hereinafter.

% figure
\begin{figure*}[t]
\begin{center}
\includegraphics[width=\textwidth]{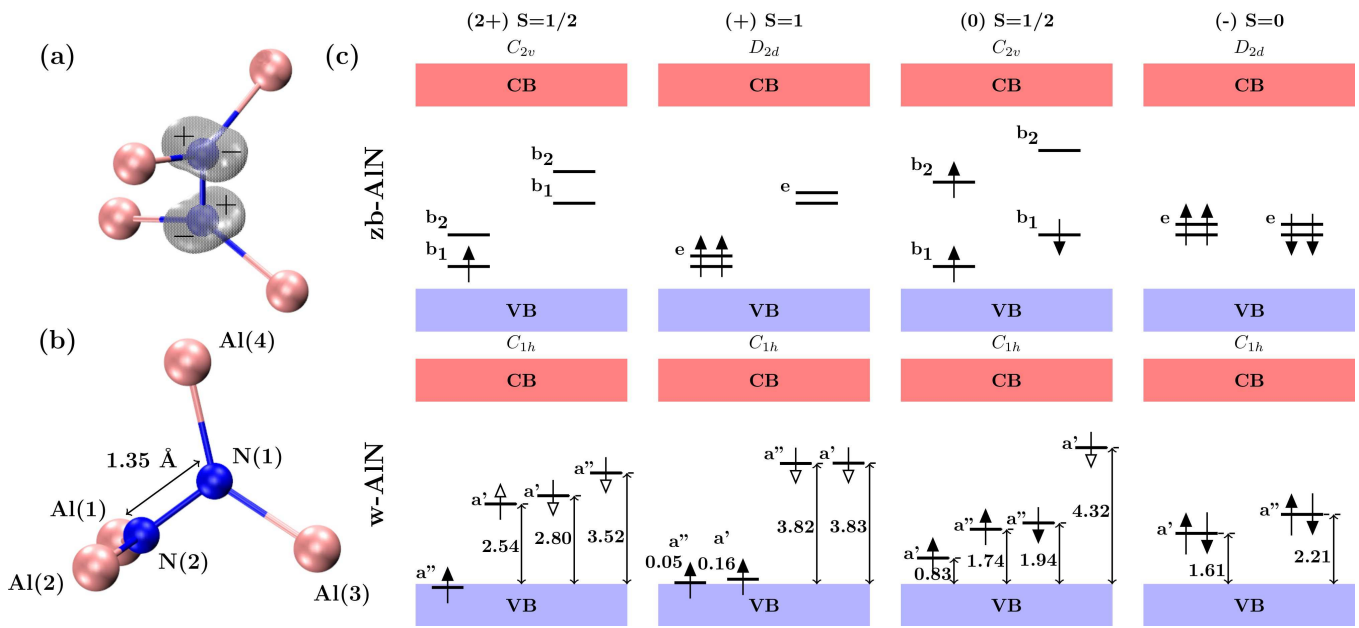}
\end{center}
\caption{(Color online) (a) The structure of core of the N split interstitial defect in the zinc-blende phase of AlN with the p orbitals forming the $e$ defect levels. (b) The close vicinity of the defect in the neutral charge state with $S=1/2$. The locations of the atoms are slightly different but the relative positions are the same in all charge states. The big pink (bright) balls are Al atoms and the small blue (dark) balls are N atoms, respectively. (c) The calculated single-electron levels in the wurtzite phase (lower part) and the expected levels in the zinc-blende phase (upper part) of the AlN respect to the valence band maximum in the ground state of the N split interstitial defect. Valence and conduction bands of the host crystal are shown as blue and red shaded regions, respectively. The arrows pointing upwards (downwards) related to the spin up (down) state. Filled (empty) arrow represents an occupied (unoccupied) level. \label{fig:band_gap}}
\end{figure*}

% figure
\begin{figure}[t]
\begin{center}
\includegraphics[scale=1.0]{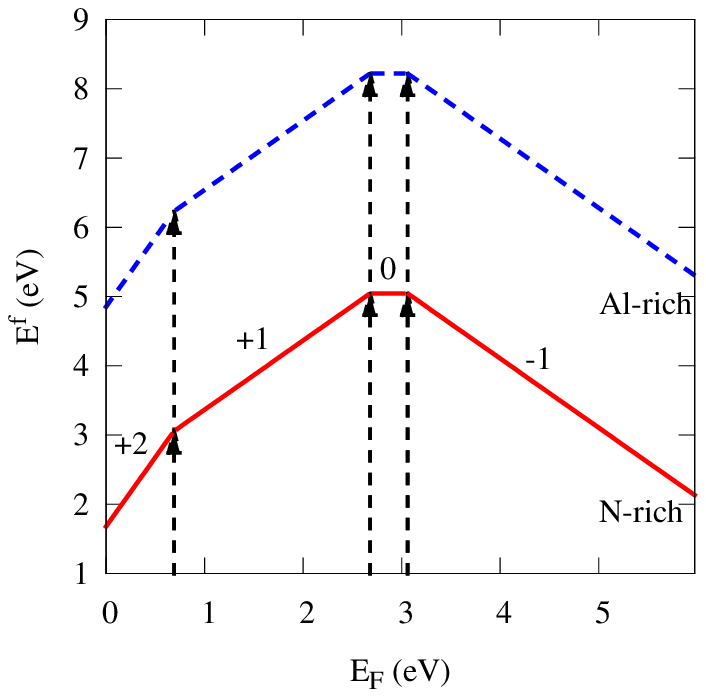}
\end{center}
\caption{(Color online) Formation energy ($E^\text{f}$) versus the Fermi level ($E_\text{F}$) position compare to the valence band maximum in different charge states of the N split interstitial defect of w-AlN in Al-rich and N-rich conditions, respectively. We consider always only the charge state with the lowest formation energy. The linear curve change its slope when the $E_\text{F}$ reach the transition level between two charge states, at $(2+/+)=0.69$ eV, at $(+/0)=2.68$ eV and at $(0/-)=3.06$ eV. \label{fig:AlN_si_formation_energy}}
\end{figure}

\section{Results and discussion}
\begin{figure}[t]
\begin{center}
\includegraphics[width=.48\textwidth]{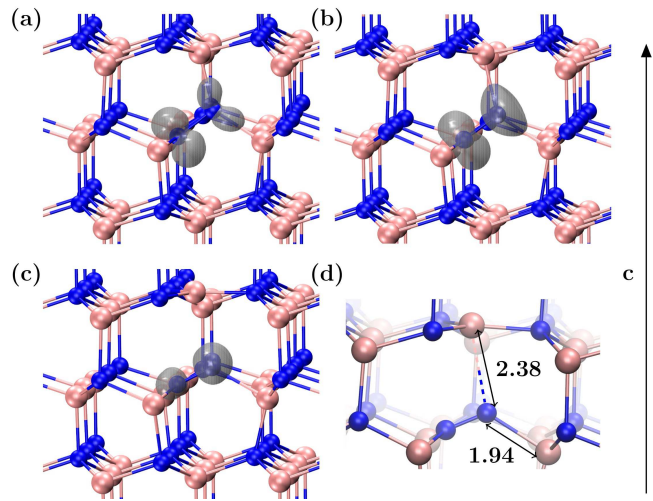}
\end{center}
\caption{(Color online) Spin polarization density plots for (a) neutral state with $S=1/2$, for (b) $(+)$ charge state with $S=1$ and (c) for $(2+)$ charge state with $S=1/2$ near the N split interstitial defect in w-AlN. (d) The close vicinity of the defect in the $(2+)$ charge state with $S=1/2$. Probably the Coulomb repulsion made the structure deformed. The bond lengths (1.94 \AA \hspace{0mm} and 2.38 \AA) are rather different inside the $C_{1h}$ symmetry plane. The big pink (bright) balls are Al atoms and the small blue (dark) balls are N atoms, respectively. (The subfigures are created with the VMD software \cite{HUMP96}) \label{fig:spin_polarization}}
\end{figure}

The structure of the pristine w-AlN has $C_{6v}$ symmetry. For the defect calculations we modify the original structure substituting one nitrogen atom with a pair of nitrogen atoms preserving the $C_{1h}$ symmetry. The distance of the two nitrogen atoms in the starting geometry for the structural optimization is set to the equilibrium nitrogen-nitrogen covalent bond length (i.e. 1.45 \AA). We use this starting geometry for all the charge and spin states. We investigate the N split interstitial defect neutral charge state in w-AlN with $S=1/2$ total spin using spin polarized calculation. We obtain four levels inside the band gap where three of them are occupied (See Fig.~\ref{fig:band_gap}c). These levels are formed by two atomic p orbitals that are localized on the two nitrogen atoms of the defect, perpendicular to the plane of the bonding orbitals (See Fig.~\ref{fig:band_gap}a). Considering the electron levels and their occupation in the neutral charge state we can expect that charge states from $(3+)$ to $(-)$ are able to form. We performed electron structure calculations in all of these charge states and we found them to be stable, except the $(3+)$. To understand better these calculated atomic single-electron levels, we considered also the zinc-blende phase of the AlN (zb-AlN), and made a group theory analysis on the levels, comparing the two structures. The results of this analysis with the single-electron levels are shown in Fig.~\ref{fig:band_gap}c with the calculated w-AlN levels and the expected zb-AlN electron levels. The zb-AlN resembles to the w-AlN, but has higher symmetry, that makes it possible to draw conclusion from the group theory analysis. Including the N split interstitial defect in the zb-AlN structure, the original $T^2_d$ space group symmetry of the bulk zb-AlN structure reduces to $D_{2d}$. The $D_{2d}$ symmetry allows to form an $e$ electron state from the two atomic p orbitals. Considering the $\text{(N-N)}_\text{N}^+$ defect in zb-AlN that has even number of electrons, the total spin of the ground state could be either $S=0$ or $S=1$. Analogue to the 1st Hund rule, the highest spin state should be lowest in energy. This implies that the $e$ state is occupied by two electrons with parallel spins, i.e. $S=1$ state. In our particular case we choose the majority spin in the spin-up channel. The structure around the core of the defect, i.e. the two nitrogen atoms and the four neighbor aluminum atoms, is very similar in zb-AlN and w-AlN. Therefore, the highly localized defect states can be similar to each other even if the crystal field stemmed from the surrounding crystal around the core of the defect is different. One can think of the surrounding crystal structure as a perturbation on the highly localized and unperturbed electron state that has the same symmetry as the core of the defect has. If this perturbation is small, the band gap states of the zb-AlN and the w-AlN will be similar. This can be seen in the Fig.~\ref{fig:band_gap}c in the case of $\text{(N-N)}_{\text{N}}^+$ in w-AlN, where two quasi degenerate states form, mimicking the $e$ state in zb-AlN. Indeed, this results in $S=1$ ground state in w-AlN. We calculate the $(+)$ charge state in two different spin state, with zero total spin and with $S=1$. We found that the $S=0$ state is about $1.17$ eV higher in energy. In the case of the $\text{(N-N)}_{\text{N}}^0$ in zb-AlN, the Jahn-Teller effect reduces the symmetry from $D_{2d}$ to $C_{2v}$. Due to this reduction, the $e$ state will split into $b_1$ and $b_2$ states. This may explain the electron levels of $\text{(N-N)}_\text{N}^0$ in w-AlN, where the energy levels split more than those in the $(+)$ charge state. At the $\text{(N-N)}_\text{N}^-$ in zb-AlN there is no Jahn-Teller effect, so $D_{2d}$ symmetry with fully occupied $e$ state is feasible. Such as in the case of the $\text{(N-N)}_\text{N}^0$ in zb-AlN, at $(2+)$ charge state the $e$-level would be occupied by a single electron which is again Jahn-Teller unstable resulting in $C_{2v}$ symmetry. The possible electron states are again $b_1$ and $b_2$. We find three levels in the band gap, but neither are occupied. The electron levels of $\text{(N-N)}_\text{N}^{2+}$ in the w-AlN are similar to $\text{(N-N)}_\text{N}^{0}$, except that one level is missing from the band gap. We find a single electron level under the VBM that is localized on the two nitrogen atoms of the defect. Probably this is the fourth defect level that is hybridized with the valence band and occupied with an $a''$ electron in the spin up channel.\\ 
\indent The N-N bond length between the two nitrogen atoms of the defect is 1.18\,\AA, 1.27\,\AA, 1.35\,\AA~and 1.47\,\AA~in the $(2+)$, $(+)$, $(0)$, $(-)$ charge states, respectively. As one can see the bond length is decreasing with the defect charge and shows close similarity with the results found in GaN \cite{Limpijumnong_PRB_2004}. In the case of the $(2+)$ charge state the neighboring geometry around the defect (i.e., the location of the Al atoms) shows significant deformation compared to the perfect crystal (See Fig.~\ref{fig:spin_polarization}d). This is probably because of the Coulomb repulsion due to the higher charge located on the defect.\\
\indent In Fig.~\ref{fig:AlN_si_formation_energy} we show the diagram of the formation energy versus the Fermi level. Comparing our results for $\text{(N-N)}_{\text{N}}$ defect in w-AlN with the results in Ref.~\onlinecite{Bardeleben_PRL_2012} for GaN and in Ref.~\onlinecite{Stampfl_PRB_2002} for zb-AlN, the same charge states are stable as a function of the Fermi level. Particularly, those charge states are interesting that have non-zero electron spins, so they might be detected by electron paramagnetic resonance techniques. According to our results, the neutral, the $(+)$ and $(2+)$ charge states possess non-zero electron spins. In material with the Fermi level located in the lower half of the band gap and under a certain level of carrier compensation, one of these charge states may appear and can be detected by EPR. Next, we further characterize these charge states with non-zero electron spins.
\\
% figure
% table
\begin{table}[h!]
\caption{Calculated principal values of the hyperfine tensor ($A_{xx}$, $A_{yy}$, $A_{zz}$) and angles ($\theta_z$, $\phi_z$) related to the $z$ component. We refer to Fig.~\ref{fig:band_gap}b for the label of the atoms. The $A_{xx}$, $A_{yy}$, $A_{zz}$ values are given in MHz, the $\theta_z$ and $\phi_z$ are determined in spherical coordinate system and are given in degree.}
\begin{ruledtabular}
\begin{tabular}{c c d d d d d}
ch. st. & \multicolumn{1}{c}{atom} & \multicolumn{1}{c}{$A_{xx}$} & \multicolumn{1}{c}{$A_{yy}$} & \multicolumn{1}{c}{$A_{zz}$} & \multicolumn{1}{c}{$\theta_z$} & \multicolumn{1}{c}{$\phi_z$}\\
\hline
\\
\parbox[t]{2mm}{\multirow{5}{*}{\rotatebox[origin=c]{90}{$(0)$ $S=1/2$}}}
& $^{14}$N(1) & -11.5 & -11.2 & 29.4 & 36.9 & 90.0 \\
& $^{14}$N(2) & -9.0 & -8.5 & 86.4 & 37.4 & 90.0 \\

& $^{27}$Al(1,2) & -25.7 & -24.5 & -25.9 & 139.4 & -108.0 \\
& $^{27}$Al(3) & 91.8 & 91.3 & 122.4 & 62.1 & 90.0 \\
& $^{27}$Al(4) & 84.5 & 84.2 & 115.3 & 9.2 & 90.0 \\
\\
\parbox[t]{2mm}{\multirow{5}{*}{\rotatebox[origin=c]{90}{$(+)$ $S=1$}}}
& $^{14}$N(1) & -13.0 & 9.2 & 37.0 & 90.0 & 0.0 \\
& $^{14}$N(2) & -11.8 & 9.2 & 36.6 & 36.4 & 90.0 \\

& $^{27}$Al(1,2) & 17.7 & 17.3 & 31.5 & 73.9 & -155.8 \\
& $^{27}$Al(3) & 20.1 & 19.6 & 34.4 & 115.5 & -90.0 \\
& $^{27}$Al(4) & 54.9 & 54.5 & 68.4 & 7.8 & 90.0 \\
\\
\parbox[t]{2mm}{\multirow{5}{*}{\rotatebox[origin=c]{90}{$(2+)$ $S=1/2$}}}
& $^{14}$N(1)  & -14.6 & -10.4 & 90.4 & 90.0 & -180.0 \\
& $^{14}$N(2)  & -20.3 & -15.6 & 29.6 & 90.0 & -180.0 \\

& $^{27}$Al(1,2)  & 39.5 & 39.0 & 64.3 & 71.1 & -154.0 \\
& $^{27}$Al(3)  &-11.6 & -9.1 &-12.0 & 90.0 & 0.0 \\
& $^{27}$Al(4)  & -1.8 &  1.0 & -1.8 & 78.5 & -90.0 \\
\end{tabular}
\end{ruledtabular}
\label{table:hyperfine}
\end{table}
We calculate the hyperfine tensors for the charge states with non-zero electron spin states for this defect. To our best knowledge, there was no such characterization concerning this defect. We collect the principal values of the hyperfine tensor for the two nitrogen atoms, $^{14}$N(1) and $^{14}$N(2) and the four nearest-neighbor aluminum atoms, $^{27}$Al(1), $^{27}$Al(2), $^{27}$Al(3) and $^{27}$Al(4) in the Table~\ref{table:hyperfine}. We refer to Fig.~\ref{fig:band_gap}b for the label of the atoms. We obtain the highest hyperfine interaction with these nuclei, except for the $(2+)$ charge state where the hyperfine tensor values on the fourth aluminum atom were much less than on the others. Probably this is related to the deformation compared to the perfect crystal in $(2+)$ charge state that we mentioned above (See Fig.~\ref{fig:spin_polarization}d). We made spin polarization density plots (electron density in the spin up channel minus the electron density in the spin down channel) focusing on the defect sites (See Fig.~\ref{fig:spin_polarization}). In the interpretation of the principal values of the hyperfine tensors, the symmetry properties of the single-electron levels may play the main role. The atomic p orbitals transform according to a different representation than all the local sigma-bond orbitals. So, they are arranged perpendicular to the plane formed by the nearest neighbor atoms (See Fig.~\ref{fig:band_gap}a) and do not take part in the bonding. The electron density related to a symmetric $a'$ level is located in the $C_{1h}$ symmetry plane (the plane of the N(1), N(2), Al(3) and Al(4) atoms in Fig.~\ref{fig:band_gap}b) and it's mostly localized on the N(2) atom and on the non-neighbor Al(3) and Al(4) atoms. Indeed, the electron density related to an antisymmetric $a''$ level is located perpendicular to the $C_{1h}$ symmetry plane and it's mostly localized on the N(1) atom and on the non-neighbor Al(1) and Al(2) atoms. In $(0)$ charge state, there are three occupied levels in the band gap where the two $a''$ levels in opposite spin channel are closer in energy and their projected wave function character are close to each other on the N(1) and N(2) atoms. This results in symmetric spin polarization density because its main contribution comes from the occupied symmetric $a'$ level. That implies higher spin polarization density and principal hyperfine value on N(2) atom than on N(1) atom and higher values on Al(3) and Al(4) atoms compare with that of on Al(1) and Al(2) atoms (See Fig.~\ref{fig:spin_polarization}a). In $(+)$ charge state there is a symmetric $a'$ and an antisymmetric $a''$ level that mimicking the $e$ state, therefore are close in energy. This implies very similar spin polarization density and principal hyperfine values either on the N(1) and N(2) sites and either on the neighbor Al sites (See Fig.~\ref{fig:spin_polarization}b). In $(2+)$ charge state the only defect level is $a''$ that implies higher spin polarization density and principal hyperfine value on N(1) atom compared with that of on N(2) atom and higher values on Al(1) and Al(2) atoms compared to that of on Al(3) and Al(4) atoms (See Fig.~\ref{fig:spin_polarization}c). We compared our hyperfine results with that of Ref.~\onlinecite{Bardeleben_PRL_2012} for GaN in both the neutral and $(2+)$ charge states. The principal hyperfine values for GaN (See Table~\ref{table:GaN_hyperfine}) may shows the same scenario as for AlN if one considers the relative values for N(1), N(2) and Ga(1), Ga(2) and Ga(3,4). Also for GaN in $(2+)$ charge state the principal hyperfine value for an N neighbor atom in the $C_{1h}$ plane ($A_\text{Ga(1)}$) disappears, presumably due the same reason, i.e. the structure deformation due to the Coulomb repulsion. Probably these similarities comes from a quite similar band gap structure and single-electron levels symmetries. 
\begin{table}[h!]
\caption{Calculated principal hyperfine tensor values $|A|$ [MHz] ($^{69}$Ga and $^{14}$N) of the $\text{(N-N)}_\text{N}$ in GaN for B\,$||$\,c from Ref.~\onlinecite{Bardeleben_PRL_2012}.}
\begin{ruledtabular}
\begin{tabular}{c c c c c c}
       & $A_{\text{N(1)}}$ & $A_{\text{N(1)}}$ & $A_{\text{Ga(1)}}$ & $A_{\text{Ga(2)}}$ & $A_{\text{Ga(3,4)}}$ \\
\hline
\\
$(0)$  & 25 & 42 & 116 & 82 & 45/50\\
$(2+)$ & 15 &  3 &     & 12 & 82   \\
\end{tabular}
\end{ruledtabular}
\label{table:GaN_hyperfine}
\end{table}

\section{Summary}

In summary, we studied the electronic structure, formation energy, transition levels and hyperfine structure of the N split interstitial defect in AlN in detail. We found four stable charge states at different Fermi levels, similarly to the same defect in GaN. We investigated the hyperfine structure of the neutral, $(+)$ and $(2+)$ charge states. We provided input for the defect identification by EPR measurement related to this defect. Under a certain level of carrier compensation the EPR may detect the neutral or (+) complex. However, using photo-EPR the (2+) charge state might be also detected. We found reasonable similarity in the hyperfine signals of $\text{(N-N)}_{\text{N}}$ defect in w-AlN when compared to those in GaN.\\

\begin{acknowledgments}
We wish to acknowledge the support from the MTA Lend\"ulet program of the Hungarian Academy of Sciences, the Swedish Energy Agency, the Swedish Foundation for Strategic Research (SSF), the Knut and Alice Wallenberg Foundation, the National Supercomputer Center in Sweden (SNIC 2013-1-331) as well as the National Information Infrastructure Development Institute in Hungary. We thank Z. Bodrog and V. Ivády for a careful reading of the manuscript.
\end{acknowledgments}

\bibliographystyle{apsrev4-1}
\bibliography{nano}

\end{document}